\documentstyle[twoside,fleqn,espcrc2]{article}

\input epsf

\font\eightrm=cmr8

\def\a{\alpha}

\def\c{\gamma}

\def\e{\epsilon}

\def\m{\mu}
\def\n{\nu}

\def\p{\pi}
\def\r{\rho}
\def\s{\sigma}
\def\t{\tau}

\def\C{\Gamma}
\def\D{\Delta}
\def\L{\Lambda}

\def\S{\Sigma}

\def\pl{\partial}
\def\ra{\rightarrow}

\def\OO{{\cal O}}
\def\BB{{\cal B}}

\def\GV{{\rm GeV}}

\def\el{{\rm e}}
\def\up{{\rm u}}
\def\dn{{\rm d}}
\def\st{{\rm s}}
\def\ch{{\rm c}}

\def\pr{{\rm p}}
\def\ne{{\rm n}}
\def\Nu{{\rm N}}
\def\ha{{\rm h}}
\def\Xx{{\rm X}}

\def\SIMQ{\mathrel{\mathop \sim_{Q^2\ra\infty}}}

\newcommand{\beq}{\begin{equation}}
\newcommand{\eeq}{\end{equation}}
\newcommand{\beqa}{\begin{eqnarray}}
\newcommand{\eeqa}{\end{eqnarray}}

\newcommand{\AmS}{{\protect\the\textfont2
  A\kern-.1667em\lower.5ex\hbox{M}\kern-.125emS}}

\title{The `Proton Spin' Effect - Theoretical Status '97}

\author{G.M. Shore \address{Department of Physics,
University of Wales Swansea, \\
Singleton Park, Swansea SA2 8PP, U.K. }
\thanks{SWAT-97/160,~hep-ph/9710367
}
\thanks{Invited review talk at QCD 97, Montpellier, July 1997}  }
 
\begin{document}
 
\begin{abstract}
The theoretical status of the `proton spin' effect is reviewed.
The conventional QCD parton model analysis of polarised DIS
is compared with a complementary approach, the composite operator 
propagator-vertex (CPV) method, each of which provides its own insight 
into the origin of the observed suppression in the first moment of $g_1^p$.
The current status of both experiment and non-perturbative calculations is
summarised. The future role of semi-inclusive DIS experiments, in both the
current and target fragmentation regions, is described.

\end{abstract}
 
\maketitle
 
\section{Introduction}

It is now nearly a decade since the EMC collaboration announced
measurements\cite{EMC} 
of the first moment of the polarised proton structure function $g_1^p$ which
in a simple parton picture could be interpreted as showing that the quarks
carry only a small fraction of the spin of the proton.
This `proton spin' problem has been the focus ever since of an extraordinary
experimental and theoretical effort to extend 
and improve the data and provide a thorough understanding of the phenomenon
in terms of QCD.

Perhaps the 1998 Montpellier conference will mark the tenth anniversary of the
EMC paper with a comprehensive review of all this work. In this ninth anniversary
review, I will instead take the opportunity to present a more individual look at 
what the `proton spin' effect means, how it can be understood in non-perturbative 
QCD, and what future experiments it suggests.

The central theme of this review is the comparision of two complementary approaches
to the description of deep inelastic scattering (DIS)-- the parton model and the
composite operator propagator-vertex (CPV) method\cite{SV1,NSV,S}. 
We will show how each provides
its own insight into the `proton spin' effect -- the first from a quark-gluon
constituent point of view and the second combining a holistic description
of the proton with non-perturbative, target-independent, QCD physics.

The status of non-perturbative calculations using QCD spectral sum rules (QSSR) 
and lattice methods is then briefly reviewed, followed by a discussion of
the current experimental situation for $g_1^p(x,Q^2)$ and the important 
unresolved questions concerning the small $x$ region.
Finally, the present and future role of semi-inclusive DIS experiments, 
both in the current and target fragmentation regions, is considered. A recent
proposal\cite{SV2} for testing the target-independent suppression mechanism
suggested by the CPV approach is described and predictions for cross
section moment ratios for $\el\Nu\ra\el{\rm h}X$ are given.

\section{The sum rule for $\C_1^p$ and the OZI rule}

The polarised structure functions are measured in polarised DIS (see Fig.~1), 
either $\m \pr \ra \m X$~(EMC,SMC) or $\el \pr \ra \el X$~(SLAC, HERMES). 
\vskip0.2cm
\centerline{
{\epsfxsize=3.5cm\epsfbox{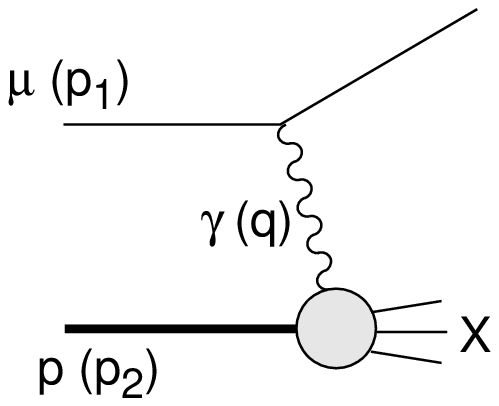}}}
\noindent{\eightrm Fig.1~~Inclusive polarised DIS scattering.}
\vskip0.2cm
\noindent $g_1^p$ is extracted from the polarisation asymmetry of the cross 
section according to:
\beq
x{d\D\s\over dx dy}~=~{Y_P\over2}{16\pi^2\a^2\over s} g_1^p(x,Q^2)~~+~~\ldots
\eeq
where $Q^2= -q^2$ and $x = Q^2/2p_2.q$ are the Bjorken variables, 
$y=Q^2/xs$ and $Y_P = (2-y)/y$.
(The dots denote terms of $O(M^2 x^2/Q^2)$,
including the second polarised structure function $g_2^p$ for which
interesting new measurements are becoming available\cite{BRULL}.)

On the theoretical side, $g_1^p$ is determined by the proton matrix element
of two electromagnetic currents carrying large spacelike momentum. 
The sum rule for the first moment $\C_1^p$ 
is derived using the twist 2, spin 1 terms in the operator product expansion 
(OPE) for the currents:
\beqa
&&J^\r(q) J^\s(-q) ~\SIMQ~
2 \e^{\r\s\n\m} {q_\n\over Q^2} ~\times \cr
&&~\Bigl[C_1^{\rm NS}(\a_s) \Bigl(A_{\m }^3 + 
{1\over\sqrt3} A_{\m }^8\Bigr)
+ {2\over3} C_1^{\rm S}(\a_s) A_{\m }^0 \Bigr] 
\eeqa
where the Wilson coefficients $C_1^{\rm NS}(\a_s)$ and $C_1^{\rm S}(\a_s)$
are now both known to $O(\a_s^3)$\cite{LAR1}.
It reads:
\beqa
&&\C^p_1(Q^2) \equiv
\int_0^1 dx~ g_1^p(x,Q^2) \cr
&&~~= {1\over12} C_1^{\rm NS} \Bigl( a^3
+ {1\over3} a^8 \Bigr) + {1\over9} C_1^{\rm S} a^0(Q^2)  
\eeqa
Here, $a^3$, $a^8$ and $a^0(Q^2)$ are the form factors in the forward 
proton matrix elements of the renormalised axial current, i.e.
\beqa
&&\langle p, s|A_{\m }^3|p, s\rangle = s_\m {1\over2} a^3 \cr
&&\langle p, s|A_{\m }^8|p, s\rangle = s_\m {1\over{2\sqrt3}} a^8 \cr
&&\langle p, s|A_{\m }^0|p, s\rangle = s_\m a^0(Q^2)
\eeqa
where $p_\m$ and $s_\m$ are the momentum and polarisation vector of the
proton. 

Because of the chiral $U_A(1)$ anomaly, $\pl^\m A_\m^0 - 2n_f Q \sim 0$,
the flavour singlet current $A_{\m }^0$ is not conserved.
It is renormalised and mixes with the topological density. Defining
the bare operators $A_{\m B}^0 = \sum \bar q \c_\m \c_5 q$
and $Q_B = {\a_s\over{8\p}} \e^{\m\n\r\s}{\rm tr} G_{\m\n} G_{\r\s}$, 
we have (for $n_f$ flavours)
\beqa
&&A_{\m }^0 = Z A_{\m B}^0 \cr
&&Q = Q_B - {1\over2n_f}(1-Z) \pl^\m A_{\m B}^0  
\eeqa
where $Z$ is a divergent renormalisation constant. The associated anomalous 
dimension $\c$ is known to 3 loops\cite{LAR2}. 
Matrix elements of $A_{\m }^0$ therefore have a non-trivial 
renormalisation group (RG) scale dependence 
governed by $\c$. In particular,
\beq
{d\over dt}a^0 = \c a^0
\eeq
where $t = \ln Q^2/\L^2$, so that the singlet axial charge $a^0(Q^2)$
is scale dependent. This is crucial to understanding the `proton spin' 
effect in QCD.

The axial charges $a^3$ and $a^8$ are known in terms of the $F$ and $D$
constants ($a^3 = F+D$, $a^8 = 3F-D$) found from neutron and hyperon 
beta decays. The interest of the sum rule centres on the flavour singlet 
axial charge $a^0(Q^2)$. In the absence of an alternative experimental
derivation of $a^0(Q^2)$, the simplest ansatz is to assume that it obeys the
OZI (Zweig) rule\footnote{The OZI limit of QCD is defined (see ref.\cite{V2})
as the truncation of full QCD in which non-planar
and quark-loop diagrams are retained, but diagrams in which the external
currents are attached to distinct quark loops (so that there are
purely gluonic intermediate states) are omitted. This is a more
accurate approximation to full QCD than either the leading large $1/N_c$ limit,
the quenched approximation (small $n_f$ at fixed $N_c$) or the
leading topological expansion ($N_c\ra\infty$ at fixed $n_f/N_c$).}, i.e. 
$a^0(Q^2) = a^8$. This gives a theoretical 
prediction for $\C_1^p$ which is known as the Ellis-Jaffe sum rule\cite{EJ}.
This is now known to be violated (see sect.~6), with $a^0(Q^2)$ strongly
suppressed relative to $a^8$. This is what is known as the `proton spin'
problem.

In fact, it is not at all surprising that the OZI rule should fail in this
case\cite{V1}. The first clue is the anomaly-induced scale dependence of 
$a^0(Q^2)$. If the OZI rule were to hold, at what scale should it be applied?
For the same reason, it is immediately clear that $a^0(Q^2)$ cannot
really measure spin. Moreover, it is known that the pseudovector
and pseudoscalar channels are linked through the Goldberger-Treiman
relations\cite{SV1}. Since large anomaly-induced OZI violations are known to be
present in the pseudoscalar channel ($U_A(1)$ problem, $\eta'$ mass, etc.)
it is natural to find them also for $a^0(Q^2)$ in the pseudovector channel.

While this immediately resolves the `proton spin' {\it problem}, clearly
we want to understand the origin of the suppression in $a^0(Q^2)$ much more
deeply. The following two sections describe two complementary approaches
to this question -- the conventional QCD parton model and the
CPV method developed in refs.\cite{SV1,NSV,S}.

\section{The QCD parton model}

In the most simple parton model, where the proton structure for large $Q^2$
DIS is described by parton distributions corresponding to free quarks only,
the polarised structure function is given by
\beq
g_1^p(x) = {1\over2} \sum_{i=1}^{n_f} ~e_i^2 ~\D q_i(x)
\eeq
where $\D q_i(x) = q_i^+(x) + \bar q_i^+(x) - q_i^-(x) - \bar q_i^-(x)$
is the difference of the distributions of quarks with helicities parallel
and antiparallel to the nucleon spin. 
It is convenient to work with the flavour non-singlet 
and singlet combinations:
\beqa
\D q^{NS}(x) &&= \sum_{i=1}^{n_f} \Bigl({e_i^2\over<e^2>} - 1\Bigr)~ 
\D q_i(x) \cr
\D q^S(x) &&= \sum_{i=1}^{n_f} ~\D q_i(x)
\eeqa
In this model, the first moment of the flavour singlet quark distribution
$\D q^S = \int_0^1 dx~\D q^S(x)$ can indeed be identified as the sum of
the helicities of the quarks. Interpreting the structure function
data {\it in this model} then leads to the conclusion that the quarks carry 
only a small fraction of the spin of the proton -- the `proton spin' problem.

However, this simple model leaves out many important features of QCD -- gluons,
RG scale dependence, the chiral $U_A(1)$ anomaly, etc. When these effects
are included, in the QCD parton model, the naive identification of
$\D q^S$ with spin no longer holds and the experimental results for
$g_1^p$ are readily accommodated.
\vskip0.2cm
\centerline{
{\epsfxsize=3.5cm\epsfbox{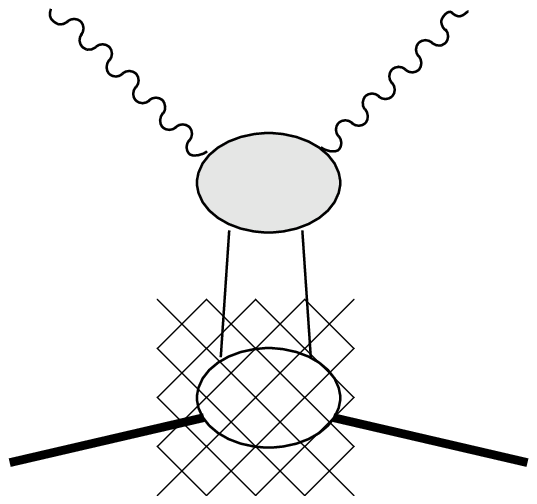}}}
\noindent{\eightrm Fig.2~~QCD parton model interpretation of DIS.
The single lines are partons, which may be quarks or gluons.}
\vskip0.2cm
The QCD parton model picture of DIS is shown in Fig.~2.
The polarised structure function is written in terms of both quark and
gluon distributions as follows:
\beqa
&&g_1^p(x,Q^2) = {1\over9}\int_x^1 {dy\over y}\Bigl[
C^{NS}\Bigl({x\over y}\Bigr)\D q^{NS}(y,t) \cr
&&~~~+ C^{S}\Bigl({x\over y}\Bigr)\D q^S(y,t) +
C^{g}\Bigl({x\over y}\Bigr)\D g(y,t) \Bigr]
\eeqa
where $C^S, C^g$ and $C^{NS}$ are perturbatively calculable functions
related to the Wilson coefficients in sect.~2 and the quark and gluon
distributions have {\it a priori} a $t=\ln Q^2/\L^2$ dependence.

The RG evolution (DGLAP) equations for these polarised distributions
are:
\beq
{d\over dt}\D q^{NS}(x,t) = {\a_s\over2\pi} \int_x^1 {dy\over y}
P_{qq}^{NS}\Bigl({x\over y}\Bigr) \D q^{NS}(y,t)
\eeq
and, abbreviating the notation of (10),
\beq
{d\over dt} \left(\matrix{\D q^S \cr \D g\cr }\right) =
{\a_s\over2\pi} \int_x^1 {dy\over y}
\left(\matrix{P_{qq}^{S} & P_{qg} \cr P_{gq} & P_{gg} \cr }\right)
\left(\matrix{\D q^S \cr \D g\cr }\right)
\eeq
showing the mixing between the singlet quark and the gluon distributions.
The splitting functions $P$ are also calculable in perturbative QCD,
their moments being related to the anomalous dimensions of the series of
increasing spin operators appearing in the OPE (2).

In this language, the first moment sum rule for $g_1^p$ reads:
\beq
\C_1^p(Q^2) = {1\over9}\Bigl[C_1^{NS}\D q^{NS}
+ C_1^S \D q^S + C_1^g \D g \Bigr]
\eeq
where $\D q^{NS}$, $\D q^S$ and $\D g$ are the first moments of the above 
distributions.
Comparing with (3), we see that the axial charge $a^0(Q^2)$ is identified 
with a linear combination of the first moments of the singlet quark
and gluon distributions. It is often, though not always, the case that
the moments of parton distributions can be identified in one-to-one
correspondence with the matrix elements of local operators. The polarised
first moments are special in that two parton distributions correspond
to the same local operator. This adds an extra subtlety to the identification.

The RG equations for the first moments of the parton distributions follow
immediately from (10,11) and depend on the matrix of anomalous dimensions
for the lowest spin, twist 2 operators. This introduces a renormalisation
scheme ambiguity. The issue of scheme dependence has been studied
thoroughly by Ball, Forte and Ridolfi\cite{BFR} and an excellent summary can 
be found in ref.\cite{BALL}. It is shown there that it is possible to 
choose a scheme known as the Adler-Bardeen or AB scheme (strictly,
a class of schemes\cite{BFR,BALL}) for which the parton
distributions satisfy the following RG equations:
\beqa
&&{d\over dt} \D q^{NS} = 0 ~~~~~~
{d\over dt} \D q^S = 0 \cr
&&{} \cr
&&{d\over dt} {\a_s\over2\pi}\D g(t) = \c \Bigl({\a_s\over2\pi}\D g(t) 
-{1\over n_f} \D q^S\Bigr)
\eeqa
with the implication $C_1^g = -n_f{\a_s\over2\p} C_1^S$.
It is then possible to make the following identifications with the axial
charges:
\beqa
a^3 &&= \D \up - \D \dn \cr
{}&{}\cr
a^8 &&= \D \up + \D \dn - 2 \D \st \cr
{}&&{}\cr
a^0(Q^2) &&= \D q^S - n_f {\a_s\over2\pi} \D g(Q^2)
\eeqa
where $\D \up = \int_0^1 dx \bigl(\D \up(x,t) + \D\bar{\up}(x,t)\bigr)$ etc.
Notice that in the AB scheme, the singlet quark distribution $\D q^S$
(which is often written as $\D \S$) is scale independent. All the scale
dependence of the axial charge $a^0(Q^2)$ is assigned to the gluon distribution
$\D g(Q^2)$.

This was the identification originally introduced for the first moments
by Altarelli and Ross\cite{AR}. We emphasise that (13) is true only
in a particular renormalisation scheme (the AB scheme) and that it is only
in this scheme that the identifications (14) hold.

In this picture, the Ellis-Jaffe sum rule follows from the assumption
that in the proton both $\D s$ and $\D g(Q^2)$ are zero.
This is equivalent to the naive OZI approximation $a^0(Q^2) = a^8$
described above. Clearly, given the RG scale dependence of $a^0(Q^2)$,
this assumption is in contradiction with QCD where the anomaly requires
$a^0(Q^2)$ to scale with the anomalous dimension $\c$.

Since neither $\D \S$ nor $\D g(Q^2)$ are currently measurable in other
processes, the parton model is unable to make a quantitative prediction
for the first moment $\C_1^p$. While the model can accomodate the
observed suppression, it cannot predict it.

An interesting conjecture, proposed in the original paper of Altarelli 
and Ross\cite{AR}, is that the observed suppression in $a^0(Q^2)$ is
due overwhelmingly to the gluon distribution $\D g(Q^2)$. If so, the
strange quark distribution $\D s \simeq 0$ in the proton and so
$\D\S \simeq a^8$. This is entirely plausible because it is the 
anomaly (which is due to the gluons and is responsible for OZI
violations) which is responsible for the scale dependence in $a^0(Q^2)$ 
and $\D g(Q^2)$ whereas (in the AB scheme) $\D\S$ is scale invariant.
The essence of this conjecture will reappear in the next section where
we describe the CPV method.

To test this conjecture, we need to find a way to measure $\D g(Q^2)$
itself, rather than the combination $a^0(Q^2)$. One possibility
(see also sect.~6) is to exploit the different scaling behaviours
of $\D q^S(x)$ and $\D g(x,Q^2)$ to distinguish their contributions in 
measurements of $g_1^p(x,Q^2)$ at different values of $Q^2$.
A second is to extract $\D g(x,Q^2)$ from processes such as open charm
production, $\c^* g \ra c \bar c$, which will be studied in various
forthcoming experiments at COMPASS, RHIC, etc.

\section{The CPV method for DIS}

This approach to sum rules in DIS was developed\cite{SV1,NSV,S} in collaboration
with S.~Narison and G.~Veneziano in a series of papers on the `proton
spin' effect. The starting point, as described above, is the use of the
OPE in the proton matrix element of two currents. This gives the standard
form for a generic structure function moment:
\beq
\int_0^1 dx~ x^{n-1} F(x;Q^2) = \sum_i C_i^n(Q^2) \langle p|\OO_i^n(0)
|p\rangle
\eeq
where $\OO_i^n$ are the set of lowest twist, spin $n$ operators in the OPE
and $C_i^n(Q^2)$ the corresponding Wilson coefficients. 
In the CPV approach, we now factorise the matrix element into the product 
of composite operator propagators and vertex functions, as illustrated in
Fig.~3.
\vskip0.2cm
\centerline{
{\epsfxsize=3.5cm\epsfbox{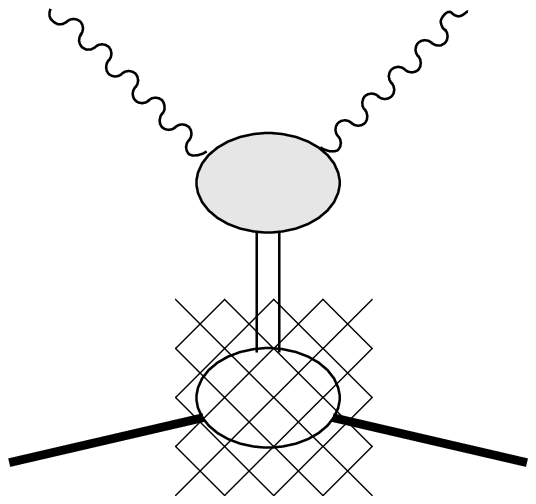}}}
\noindent{\eightrm Fig.3~~CPV description of DIS. The double line denotes
the composite operator propogator and the lower blob the `1PI' vertex.}
\vskip0.2cm
To do this, we first select a set of composite operators $\tilde {\OO}_i$
appropriate to the physical situation and define vertices 
$\C_{\tilde{\OO}_i pp}$ as `1PI' with respect to this set.
Technically, this is achieved by introducing sources for these operators 
in the QCD generating functional, then performing a Legendre transform
to obtain an effective action $\C[\tilde{\OO}_i]$. The 1PI vertices are the 
functional derivatives of $\C[\tilde{\OO}_i]$.
The generic structure function sum rule (15) then takes the form
\beqa
&&\int_0^1dx~x^{n-1}~F(x,Q^2) =  \cr
&&~~~\sum_i \sum_j C_j^{(n)}(Q^2) \langle0|T~\OO_j^{(n)} ~
\tilde{\OO}_i |0\rangle \C_{\tilde{\OO}_i pp}
\eeqa

This decomposition splits the structure function into three pieces -- first,
the Wilson coefficients $C_j^{(n)}(Q^2)$ which control the $Q^2$ dependence 
and can be calculated in perturbative QCD; second, non-perturbative but
{\it target-independent} QCD correlation functions (composite operator
propagators) $\langle0|T~\OO_j^{(n)} ~\tilde{\OO}_i |0\rangle$; and third,
non-perturbative, target-dependent vertex functions $\C_{\tilde{\OO}_i pp}$
describing the coupling of the target proton to the composite operators 
of interest.
The vertex functions cannot be calculated directly from first principles.
They encode the information on the nature of the proton state and play an 
analogous role to the parton distributions in the more conventional
parton picture. 

As emphasised in refs.\cite{SV1,NSV,S}, it is important to recognise that this
decomposition of the matrix elements into products of propagators
and proper vertices is {\it exact}, independent of the choice of
the set of operators $\tilde{\OO}_i$. In particular, it is not necessary
for $\tilde{\OO}_i$ to be in any sense a complete set. All that happens if a 
different choice is made is that the vertices $\C_{\tilde{\OO}_i pp}$
themselves change, becoming `1PI' with respect to a different
set of composite fields. Of course, while any set of $\tilde{\OO}_i$ may be
chosen, some will be more convenient than others. Clearly, the set 
of operators should be as small as possible while still capturing the
essential physics (i.e.~they should encompass the relevant degrees of
freedom) and indeed a good choice can result in vertices $\C_{\tilde{\OO}_i pp}$
which are both RG invariant and closely related to low energy physical 
couplings, such as $g_{\p NN}$ or $g_{\p\c\c}$\cite{SV1}. In this case, 
(16) provides a rigorous relation between high $Q^2$ DIS and low-energy 
meson-nucleon scattering. 

For the first moment sum rule for $g_1^p$, it is most convenient to use
the chiral anomaly immediately to re-express $a^0(Q^2)$ 
in terms of the forward matrix element of the topological density $Q$, i.e.
\beq
a^0(Q^2) ~=~{1\over 2M} 2n_f \langle p|Q|p\rangle
\eeq

Our set of operators $\tilde{\OO}_i$ is then chosen to be the renormalised 
flavour singlet pseudoscalars $Q$ and $\Phi_{5}$ where, up to a crucial 
normalisation factor, the corresponding bare operator is 
$\Phi_{5B} = \sum \bar q \c_5 q$.
This normalisation factor is chosen such that in the absence of the 
anomaly, or more precisely in the OZI limit of QCD (see footnote 3), 
$\Phi_{5}$ would have the correct normalisation to couple with unit decay 
constant to the $U(1)$ Goldstone boson which would exist in this limit. 
This also ensures that the vertex is RG scale independent\cite{SV1}.
We then have
\beqa
&&\C_{1~sing}^p ={1\over9} {1\over2M} 2n_f
C_1^{\rm S}(\a_s) ~\times \cr 
&&\biggl[\langle 0|T~Q~ Q|0\rangle \C_{Qpp}
+\langle 0|T~ Q~ \Phi_{5}|0\rangle \C_{\Phi_5 pp} \biggr]
\eeqa
where the propagators are at zero momentum and the vertices 
are 1PI wrt $Q$ and $\Phi_{5}$ only. This is illustrated in Fig.~4.
\vskip0.2cm
\centerline{
{\epsfxsize=7cm\epsfbox{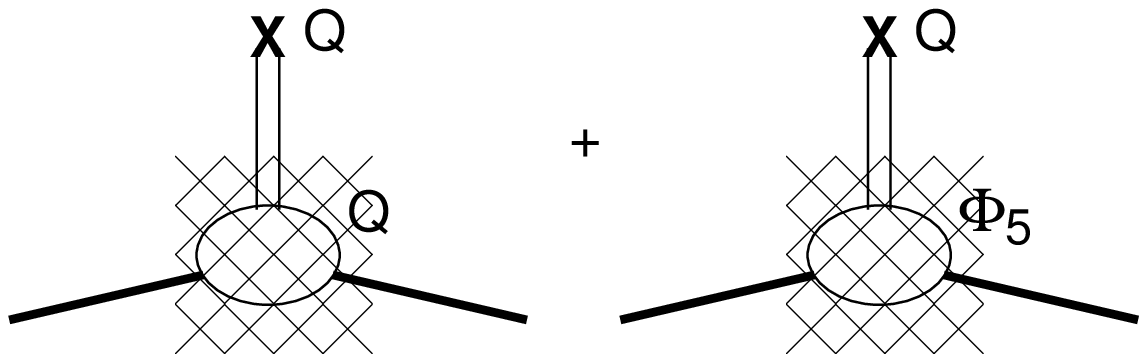}}}
\vskip0.2cm
\noindent{\eightrm Fig.4~~CPV decomposition of the matrix element 
$\langle p|Q|p\rangle$.}
\vskip0.2cm
The composite operator propagator in the first term is the zero-momentum
limit of the QCD topological susceptibility $\chi(k^2)$, viz.
\beq
\chi(k^2) = \int dx e^{ik.x} i\langle 0|T~ Q(x)~Q(0)|0\rangle
\eeq
The anomalous chiral Ward identities show that $\chi(0)$ vanishes 
for QCD with massless quarks, in contrast to pure Yang-Mills theory where
$\chi(0)$ is non-zero. Furthermore, it can be shown\cite{SV1} that the  
propagator $\langle 0|T~ Q~\Phi_{5}|0\rangle$ at zero momentum
is simply the square root of the first moment of the topological
susceptibility. We therefore find:
\beq
\C_{1~sing}^p ~=~ {1\over9} {1\over 2M} 2n_f~
C_1^{\rm S}(\a_s) ~\sqrt{\chi^{\prime}(0)} ~\C_{\Phi_{5} pp}
\eeq
The quantity $\sqrt{\chi^\prime(0)}$ is
not RG invariant and scales with the anomalous dimension $\c$, i.e.
\beq
{d\over dt} \sqrt{\chi'(0)} = \c \sqrt{\chi'(0)}
\eeq
On the other hand, the proper vertex has been chosen specifically
so as to be RG invariant. The renormalisation group properties of this
decomposition are crucial to the resolution proposed in ref.\cite{SV1,NSV} 
of the `proton spin' problem.
 
The proposal of ref.\cite{SV1,NSV} is that we should expect the source of OZI 
violations to lie in the RG non-invariant, and therefore anomaly-sensitive, 
terms, i.e. in $\chi^{\prime}(0)$.
Notice that we are using RG non-invariance, i.e.~dependence 
on the anomalous dimension $\c$, merely as an indicator of which quantities
are sensitive to the anomaly and therefore likely to show OZI violations.
Since the anomalous suppression in $\C_1^p$ is assigned to the composite 
operator propagator rather than the proper vertex, the suppression is a 
{\it target independent} property of QCD related to the chiral anomaly, 
not a special property of the proton structure\footnote{
Other models which predict target independence exist in the literature.
In ref.\cite{BALLG} it is suggested that the suppression in $a^0(Q^2)$
may be due directly to non-perturbative effects in $\c$ at low scales.
In the model of ref.\cite{FRIT}, the axial charge is related
to the nucleon couplings of the pseudovector mesons, which are relatively
uncertain experimentally, rather than using the Goldberger-Treiman 
relations to compare with the pseudoscalar sector.}.
  
To convert this into a quantitative prediction we use the OZI 
approximation for the vertex  $\C_{\Phi_{5} pp}$ and a QCD spectral sum 
rule estimate of the first moment of the topological 
susceptibility. We find\cite{NSV}, for $n_f=3$,  
\beq
\sqrt{\chi^\prime(0)}\Big|_{Q^2=10 GeV^2}
= 23.2 \pm 2.4~{\rm MeV}
\eeq
This is a suppression of approximately a factor $0.6$ relative to the 
OZI value $f_\pi /\sqrt6$. 

Our final result, in the chiral limit, is then
\beq
a^0(Q^2=10\GV^2) ~=~ 0.35 \pm 0.05
\eeq
from which we deduce
\beq
\C_1^p\Big|_{Q^2=10\GV^2} ~=~ 0.143 \pm 0.005
\eeq
This is to be compared with the Ellis-Jaffe (OZI) prediction of $a^0 = 0.58$.

The complementary nature of the QCD parton model and CPV methods is now
clear. Both involve at present incalculable non-perturbative functions
describing the proton state -- the quark and gluon distributions in the 
parton picture and the 1PI vertices in the CPV method. Both exhibit a degree
of universality -- the same parton distributions may be used in different
QCD processes such as DIS or hadron-hadron collisions, while the vertices
(when they can be identified with low-energy couplings) also provide a link
between high $Q^2$ DIS and soft meson-nucleon interactions.

One of the main advantages of the CPV method is that some non-perturbative 
information which is generic to QCD, i.e.~independent of the target, is
factored off into the composite operator propagator. This allows us to 
distinguish between non-perturbative mechanisms which are generic to all
QCD processes and those which are specific to a particular target.
As explained above, our contention is that the anomalous suppression in the 
first moment of $g_1^p$ is of the first, target-independent, type.
This target-independence conjecture can in principle be tested by
DIS with non-nucleon targets. This may effectively be realised in
semi-inclusive DIS (see sect.~7 and ref.\cite{SV2}).

Both the parton and CPV methods allow a natural conjecture in which the 
origin of this suppression is attributed to `glue' -- either through a 
large polarised gluon distribution $\D g(Q^2)$ in the parton description or 
due to an anomalous suppression of the first moment of the topological 
susceptibility $\sqrt{\chi'(0)}$ in the CPV description.
These conjectures are based on assumptions that the appropriate RG invariant
quantities, $\D \S$ or $\C_{\Phi_5 pp}$, obey the OZI rule.
The motivation for this is particularly strong in the CPV case, where it is
supported by a range of evidence from low-energy meson phenomenology
in the $U_A(1)$ channel.

The two approaches therefore provide related, but complementary, insights 
into the nature of the `proton spin' effect. Perhaps the most important
message of this review is that both insights are needed and both methods 
have a full part to play in understanding this intriguing and subtle phenomenon.

\section{Non-perturbative results -- QSSR and lattice}

The challenge to non-perturbative calculational methods in QCD is therefore
to evaluate $a^0(Q^2)$, either directly through the matrix element
$\langle p|A_\m^0|p\rangle$ or, using the anomaly, from $\langle p|Q|p\rangle$.
Alternatively, if we accept the conjecture proposed above, we can deduce
$a^0(Q^2)$ from a calculation of $\chi'(0)$. In any case, the topological
susceptibility is a fundamental correlation function in QCD of great importance
in a variety of contexts and deserves to be studied in its own right.

These non-perturbative calculations can be performed either using lattice
gauge theory or the method of QCD spectral sum rules (QSSR).
It must be stressed that to be meaningful, these calculations must be performed 
in full QCD, or at the very least in an approximation that incorporates
chiral $U_A(1)$ breaking by the anomaly. The occurrence of massless
singularities is quite different in truncations such as large $N_c$
or the quenched approximation (see footnote 3) where the anomaly is not
properly included, due to the erroneous appearance of a $U_A(1)$ Goldstone 
boson.

Lattice calculations of $a^0(Q^2)$ from $\langle p|A_\m^0|p\rangle$ have been
performed by several authors and are reviewed in ref.\cite{LIU1}. 
Since only quenched configurations have been used, the results cannot
give the true $a^0(Q^2)$. However, a major step towards the true answer can be
made by including explicitly OZI-violating `disconnected' diagrams with 
purely gluonic intermediate states as well as the OZI respecting 
`connected' diagrams (see Fig.~5).
\vskip0.2cm
\centerline{
{\epsfxsize=7cm\epsfbox{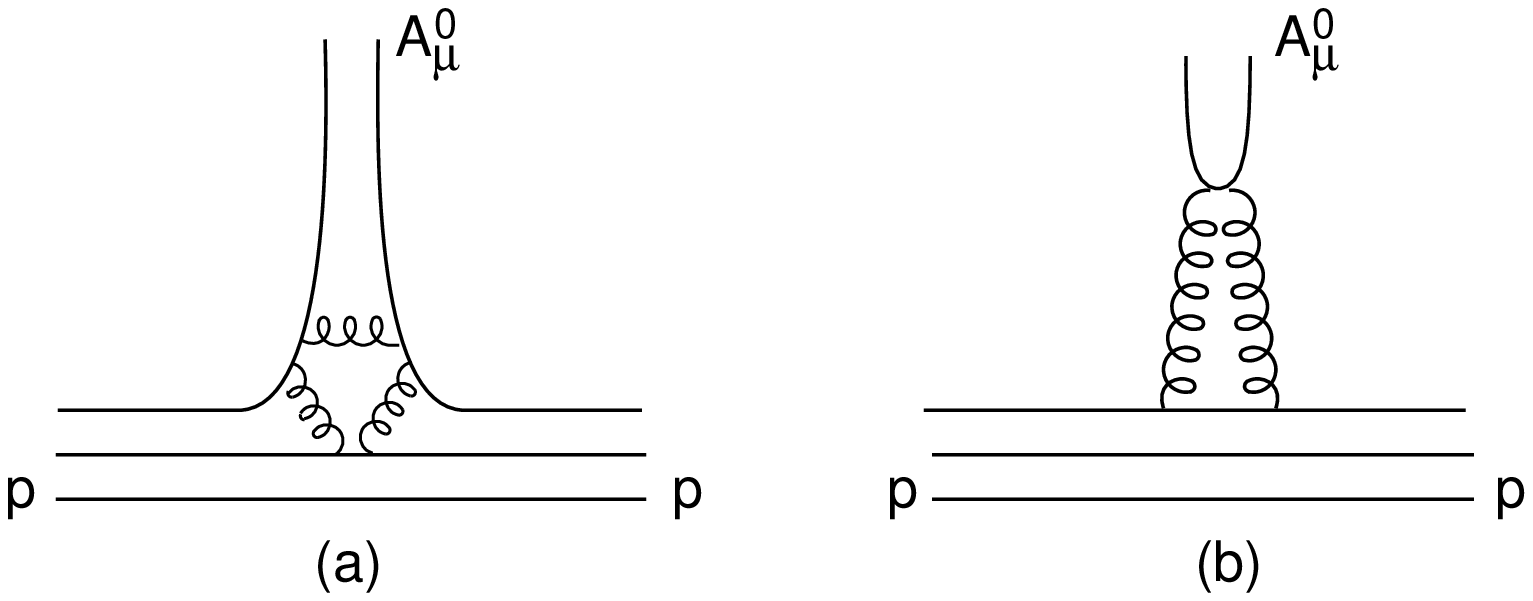}}}
\noindent{\eightrm Fig.5~~`Connected' (a) and `disconnected' (b) contributions
to the matrix element $\langle p|A_\m^0|p\rangle$.}
\vskip0.2cm
Using this method, Dong, Laga\"e and Liu \cite{DLL,LIU2} find
$a^0 = 0.25 \pm 0.12$, and confirm that the contribution from the
connected diagrams ($a^0_{conn} = 0.62\pm0.09$) reproduces the OZI expectation.
It is not clear, however, how much this result will change if non-quenched
gauge configurations including the effects of dynamical fermions are used.

For $\langle p|Q|p\rangle$, so far only quenched calculations have been
performed\cite{LIU1}, although the Pisa group\cite{GIAC} are currently running
simulations in full QCD. Existing calculations of proton matrix elements 
using QSSR have also not fully incorporated OZI breaking.

The first moment of the topological susceptibility, $\chi'(0)$, has been
evaluated using QSSR in ref.\cite{NSV} and the result is quoted above in (22).
This shows $\sqrt{\chi'(0)}$ to be suppressed to around 0.6
of the OZI value. The method and calculational details are described 
in full in \cite{NSV}.

The application of QSSR to the $U_A(1)$ sector of QCD has been criticised
repeatedly by Ioffe (see e.g.~\cite{IOFFE}) on two grounds:~ (i)~that there
are important neglected contributions from `instantons', i.e.~higher dimensional
condensates, and ~(ii)~ that when the strange quark mass is included, QSSR
results for current-current correlators show quite unrealistic $SU(3)$ breaking.
Neither criticism is valid, and both will be refuted in detail in a forthcoming
paper\cite{NSV2}. In fact, the stabilisation scale in the calculation of 
ref.\cite{NSV}, viz.~$\t^{-1} \sim 2\GV^2$, is sufficiently big for
higher dimensional condensates to be suppressed, and indeed the calculation
does display the hierarchy of gluonium to light meson hadronic scales
anticipated by ref.\cite{NSVZ}. Ioffe's second criticism is based on a 
calculation\cite{IOFFE} where radiative corrections are not properly implemented,
giving a false result. The correct results are given in ref.\cite{NSV2},
where we extend our previous analysis systematically beyond the chiral limit
using a new set of generalised Goldberger-Treiman relations.

$\chi'(0)$ is a particularly difficult correlation function to calculate on the
lattice, requiring algorithms that implement topologically non-trivial
configurations in a sufficiently fast and efficient way.
However, very preliminary results from the Pisa group\cite{GIAC} of calculations
in full QCD with dynamical quarks indicate a value of 
$\sqrt{\chi'(0)} \simeq 16 \pm 3 {\rm MeV}$. Given the preliminary nature
of the lattice simulations, the agreement with the QSSR result (22) is
encouraging and further results from the lattice are eagerly awaited.

\section{Experiment and the small $x$ region}

The most recent published results from the SMC collaboration on $g_1^p$
are given in ref.\cite{SMC1}. For the first moment, SMC quote:
\beq
\C_1^p\Big|_{Q^2=10\GV^2} = 0.136 \pm 0.013 \pm 0.009 \pm 0.005
\eeq
where the last error is a theoretical one, related to the evolution
of the measured data to a standard $Q^2$. This implies
\beq
a^0(Q^2=10\GV^2) ~=~ 0.28 \pm 0.16
\eeq
This is to be compared with the prediction (23,24) obtained in sect.~4.

Although this agreement is very promising and suggests strongly
(see also below) that the explanation of the `proton spin' effect in 
terms of a suppression in the topological susceptibility is correct,
there is an important uncertainty in the presentation of this data.
In fact, SMC only take measurements in the region $x>0.003$. The
contribution of the small $x$ region to $\C_1^p$ is estimated in
ref.\cite{SMC1}
by a dubious Regge extrapolation, which gives only a very small addition of
$0.0042\pm0.0016$. Ball, Forte and Ridolfi\cite{BFR} (see also \cite{ABFR})
have proposed an alternative small $x$ extrapolation, using the data to
fit the parton densities at small $x$ at a low $Q^2$ scale, then using 
the perturbative QCD evolution equations to deduce the form of 
$g_1^p(x, Q^2=10\GV^2)$. Examples of their fits are shown in Fig.~6.
\vskip0.2cm
\centerline{
{\epsfxsize=7cm\epsfbox{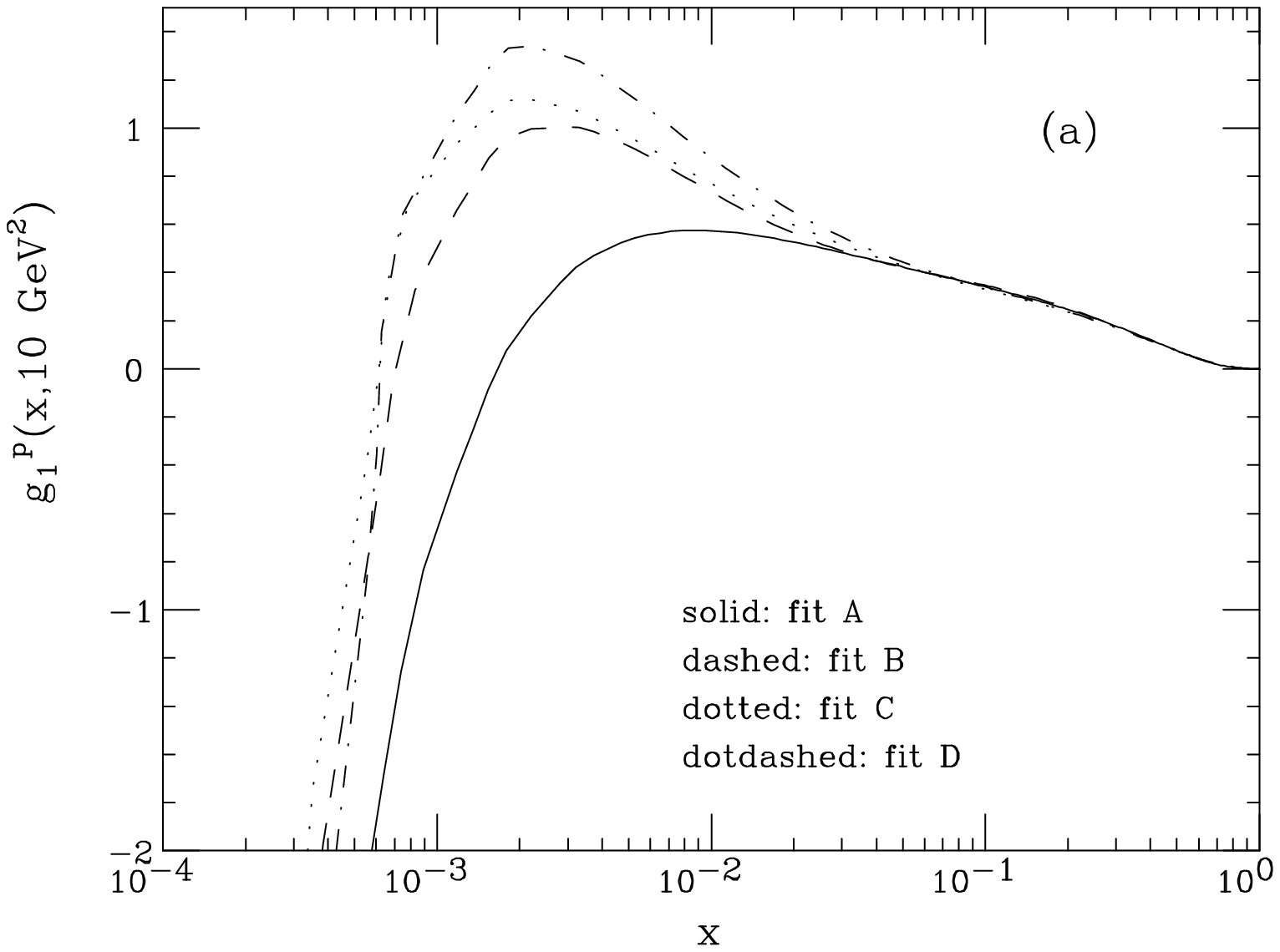}}}
\noindent{\eightrm Fig.6~~QCD evolved $g_1^p(x, Q^2=10\GV^2)$ at small $x$.}
\vskip0.2cm
The important feature of the evolution equations is that while
$\D\S(x,Q^2)$ falls with increasing $Q^2$, the polarised gluon distribution
$\D g(x,Q^2)$ rises. The net effect is that $g_1^p(x,Q^2)$ is driven strongly
negative at $Q^2=10\GV^2$ for sufficiently small $x$.
This gives a potentially large negative contribution to the first moment
$\C_1^p$, but with relatively large errors.

This summer, SMC\cite{SMC2,SMC3} have released new, still preliminary, data
on $g_1^p$ based on the 1996 run. See Fig.~7.
\vskip0.2cm
\centerline{
{\epsfxsize=6.5cm\epsfbox{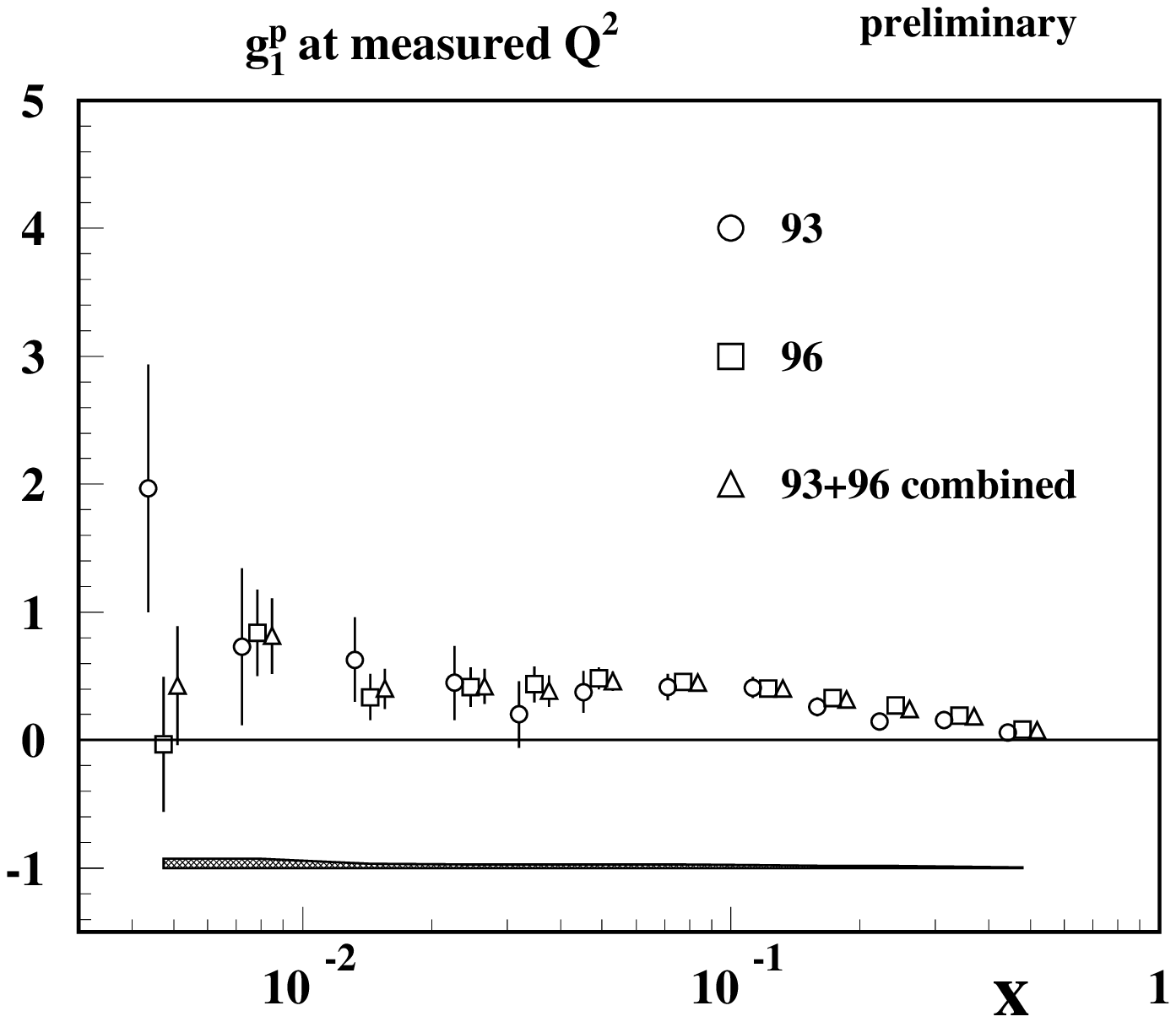}}}
\noindent{\eightrm Fig.7~~Preliminary SMC data for $g_1^p$}
\vskip0.2cm
It shows two significant features compared with the older data -- the 
smallest $x$ point has dropped appreciably, while there is a small rise in the
medium $x$ data points. 
\vskip0.2cm
\centerline{
{\epsfxsize=6cm\epsfbox{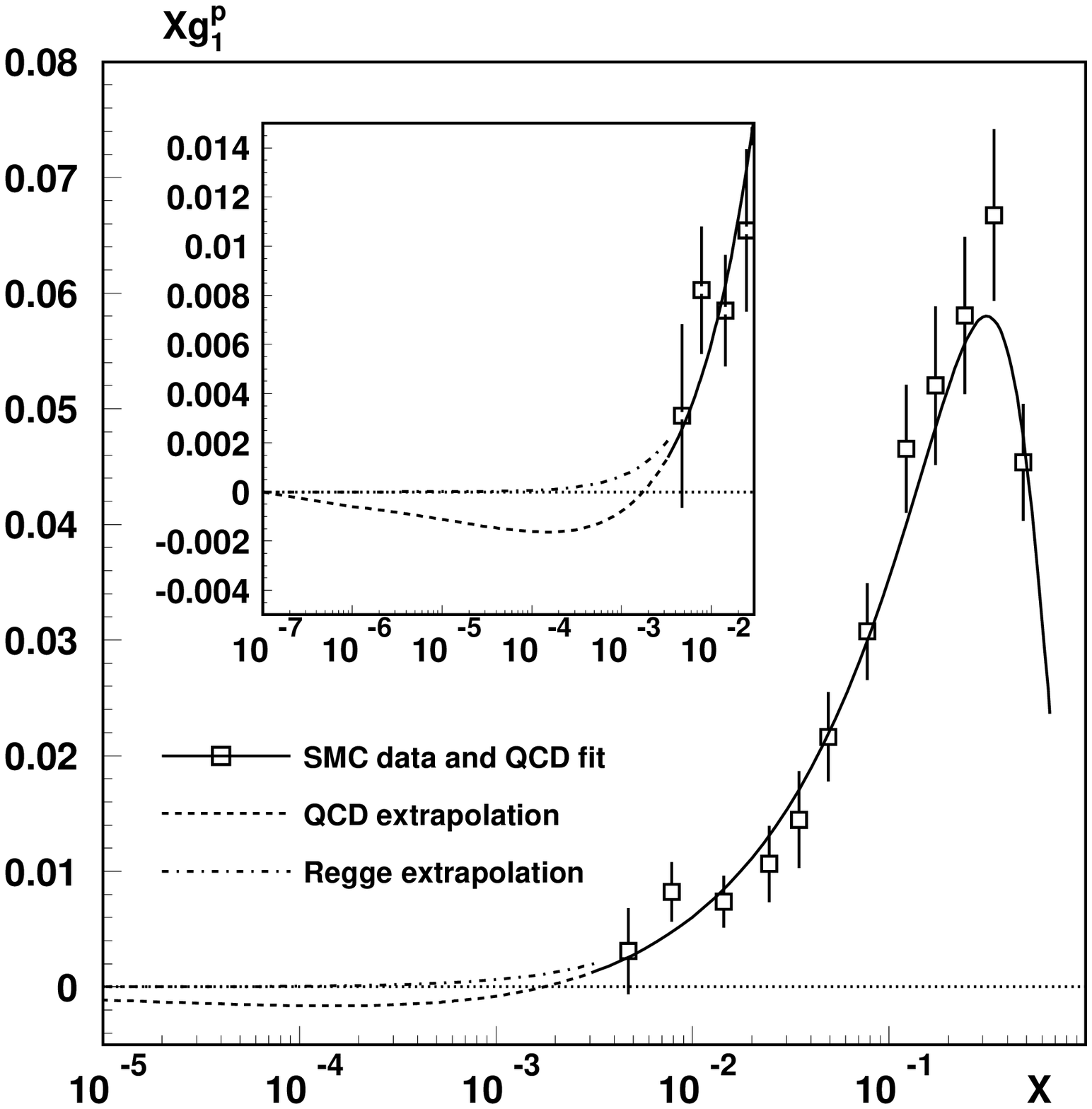}}}
\vskip0.4cm
\noindent{\eightrm Fig.8~~Preliminary SMC data for $xg_1^p$ including
`Regge' and `QCD' small $x$ extrapolations.}
\vskip0.2cm
In presenting these results (see Fig.~8), SMC have given fits to the
small $x$ region using both the `Regge' and `QCD' extrapolations. They
quote the following results. For the measured range (including an 
uncontroversial $x>0.7$ extrapolation):
\beqa
&&\int_{0.003}^1 dx~g_1^p(x;Q^2=10\GV^2) \cr
&&~~~~~~=~ 0.146 \pm 0.006 \pm 0.009 \pm 0.005  
\eeqa
while with small $x$ extrapolations:
\beq
\C_1^p\Big|_{Q^2=10\GV^2}^{Regge} = 0.149 \pm 0.006 \pm 0.009 \pm 0.005
\eeq
\beq
a^0(Q^2=10\GV^2){}^{Regge} = 0.41 \pm 0.11
\eeq
and
\beq
\C_1^p\Big|_{Q^2=10\GV^2}^{QCD} = 0.135 \pm 0.006 \pm 0.009 \pm 0.011 
\eeq
\beq
a^0(Q^2=10\GV^2){}^{QCD} = 0.27 \pm 0.15
\eeq
Notice that these results are for $g_1^p$ taken from proton data alone.
The values of $\C_1^p$ and $a^0$ for $g_1^p$ taken from combined proton
and deuteron data are systematically lower\cite{SMC2,SMC3}.

The small $x$ region is therefore an important challenge to experimentalists.
Resolving the above uncertainty in $\C_1^p$ and $a^0$ will enable a
rigorous test of the CPV conjecture and the link with the topological
susceptibility. In parton terms, measuring the small $x$ region 
at different $Q^2$ will, through analysis of the RG scaling behaviour,
enable the gluon distribution $\D g(x,Q^2)$ to be isolated\cite{BFR,BFRH}
accurately.
These will be important tasks for future experiments at HERA with a polarised
$\pr$ beam.

\section{Semi-inclusive DIS}

As well as pushing further into the small $x$ frontier, an increasingly
important role in the study of polarisation and the proton structure
will be played in future by semi-inclusive processes. We have already 
mentioned the importance of open charm production in isolating the gluon 
distribution $\D g(x,Q^2)$. In this section, we describe what may be learnt 
from semi-inclusive DIS, both in the current and target fragmentation regions.
\vskip0.2cm
\centerline{
{\epsfxsize=2.8cm\epsfbox{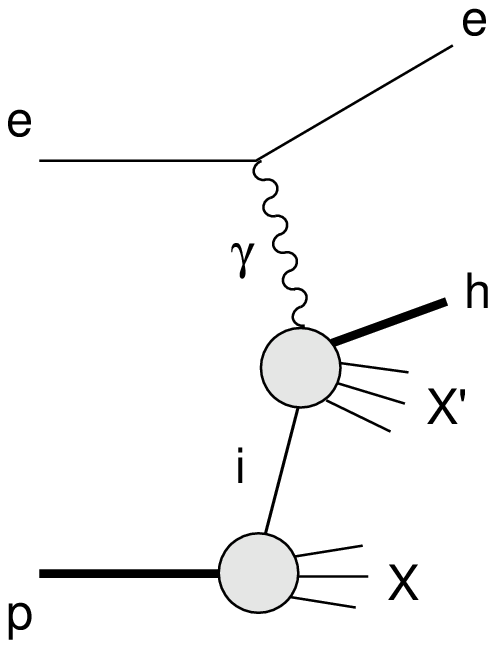}}}
\noindent{\eightrm Fig.9~~Semi-inclusive DIS: current fragmentation region.}
\vskip0.0cm
\centerline{
{\epsfxsize=2.8cm\epsfbox{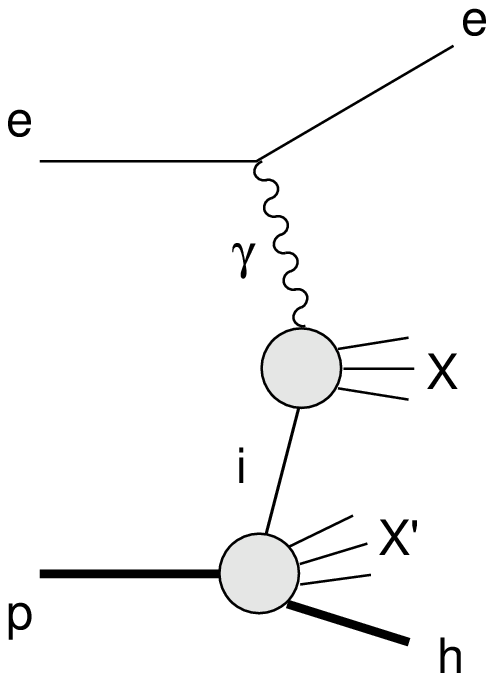}}}
\noindent{\eightrm Fig.10~~Semi-inclusive DIS: target fragmentation region.}
\vskip0.2cm

The two distinct contributions to the semi-inclusive DIS reaction 
$\el \Nu \ra \el \ha X$ from the current
and target fragmentation regions are shown in Figs.~9 and 10.
The current fragmentation events are described by parton fragmentation
functions $D_i^h({\tilde z\over 1-x},Q^2)$, where $i$ denotes the parton, while 
the target fragmentation events are described by fracture functions\cite{TV}
$M_i^{hN}(x,\tilde z,Q^2)$ representing the joint probability distribution for
producing a parton with momentum fraction $x$ and a detected hadron $\ha$ 
(with momentum $p_2^{\prime}$) carrying energy fraction $\tilde z$ from a 
nucleon $\Nu$.

The lowest order cross section for polarised semi-inclusive DIS is:
\beqa
&&x~{d\D\s \over dx dy d\tilde z} = {Y_P\over2} {4\pi\a^2 \over s} \sum_i 
e_i^2 \biggl[\D M_i^{hN}(x,\tilde z,Q^2) \cr
&&~~~~~+ {1\over 1-x} \D q_i(x,Q^2) D_i^h\Bigl({\tilde z\over 1-x}, Q^2\Bigr)
\biggr]
\eeqa
where $\tilde z = E_h/E_N$\cite{DEF}.
The notation is slightly different from sect.~3. Here, $\D q_i(x)$
refers to quarks and antiquarks separately and a sum over both is implied.
$\D M_i^{hN}(x,\tilde z,Q^2)$ is the polarisation asymmetry of the 
fracture function. The NLO corrections to (32) are given in ref.\cite{DEF}.

Quite different physics emerges from the two regions. A programme
of semi-inclusive DIS in the current fragmentation region has already been 
carried out by SMC\cite{SMC4} and will be pursued by HERMES and COMPASS.
The particular interest is that it allows the distributions for quarks
and antiquarks to be separated, giving information on the polarised
`valence' and `sea' distributions defined as 
$\D q_i^v(x) = \D q_i(x) - \D \bar q_i(x)$
and $\D q_i^{sea}(x) = 2\D \bar q_i(x)$.
These are found by comparing cross section asymmetries for positive and
negative charged hadrons $h$, with the assumption that the fragmentation
functions $D_i^h$ in (32) satisfy isospin and charge conjugation symmetry
and are helicity independent\cite{F}.
The first results for the moments $\D\up^v$ and $\D\dn^v$ are given
in ref.\cite{SMC4}.

Recently, a new proposal to exploit semi-inclusive DIS in the target
fragmentation region to elucidate the `proton spin' effect has been 
presented\cite{SV2}. The idea is to test the `target independence' conjecture
suggested by the CPV method by using semi-inclusive DIS in effect to make
measurements of the polarised structure functions of other hadronic targets
besides the proton and neutron.

The basic conjecture of ref.\cite{NSV} is that for any hadron, the singlet
axial charge in (3) can be substituted by its OZI value multiplied by a 
universal (target-independent) suppression factor $s(Q^2)$ determined, up to 
radiative corrections, by the anomalous suppression of the first moment
of the topological susceptibility $\sqrt{\chi'(0)}$. For example,
for a hadron containing only $\up$ and $\dn$ quarks, the OZI relation
is simply $a^0 = a^8$, so we predict:
\beq
\C_1 = {1\over12} C_1^{\rm NS} \Bigl( a^3 + {1\over3}(1 + 4s) a^8 \Bigr)
\eeq
where
\beq
s(Q^2) = {C_1^{\rm S}(\a_s) \over C_1^{\rm NS}(\a_s)}~{a^0(Q^2)\over a^8}
\eeq
Since $s$ is target independent, we can use the value measured for the proton 
to deduce $\C_1$ for any other hadron simply from the flavour 
non-singlet axial charges.
From our spectral sum rule estimate of $\sqrt{\chi'(0)}$, we find 
$s\sim 0.66$ at $Q^2=10{\rm GeV}^2$, while the central value of the SMC 
result (26) gives $s\sim 0.55$. 

The non-singlet axial charges for a hadron $\BB$ are given by the matrix
elements of the flavour octet axial currents, so can be factorised
into products of $SU(3)$ Clebsch-Gordon coefficients times reduced matrix
elements. Together with the target independence conjecture, this allows
predictions to be made for ratios of the first moments of the
polarised structure functions $g_1^{\BB}$ for different $\BB$.
Some of the most intriguing are:
\beq
\C_1^p / \C_1^n ~~=~~ {2s -1 -3(2s+1)F/D \over 2s +2 -6sF/D} 
\eeq
\beq
\C_1^{\D^{++}} / \C_1^{\D^-} ~~=~~ \C_1^{\S_c^{++}} / \C_1^{\S_c^0} ~~=~~
 {2s+2 \over 2s-1} 
\eeq
where $\S_c^{++}$ ~($\S_c^0$) is the state with valence quarks $\up\up\ch$~
($\dn\dn\ch$).
The results for $\D$ and $\S_c$ are particularly striking because of the 
$2s-1$ denominator factor, which is very small for the range of $s$ favoured by
experiment. These examples therefore show spectacular deviations from the 
valence quark counting (OZI) expectations, which would give the ratio 4.

The proposal of ref.\cite{SV2} is that these ratios can be realised in 
semi-inclusive DIS in a kinematical region where the detected hadron
$\ha$ (a pion or {\rm D} meson in these examples) carries a large target
energy fraction, i.e.~$\tilde z$ approaching 1. While this process is most rigorously
described in terms of fracture functions, it can be pictured as the single
Regge exchange shown in Fig.~11. The fracture function description,
which utilises the recently introduced extended fracture functions of
ref.\cite{GTV,T}, may be found in ref.\cite{SV2}.
\vskip0.0cm
\centerline{
{\epsfxsize=3cm\epsfbox{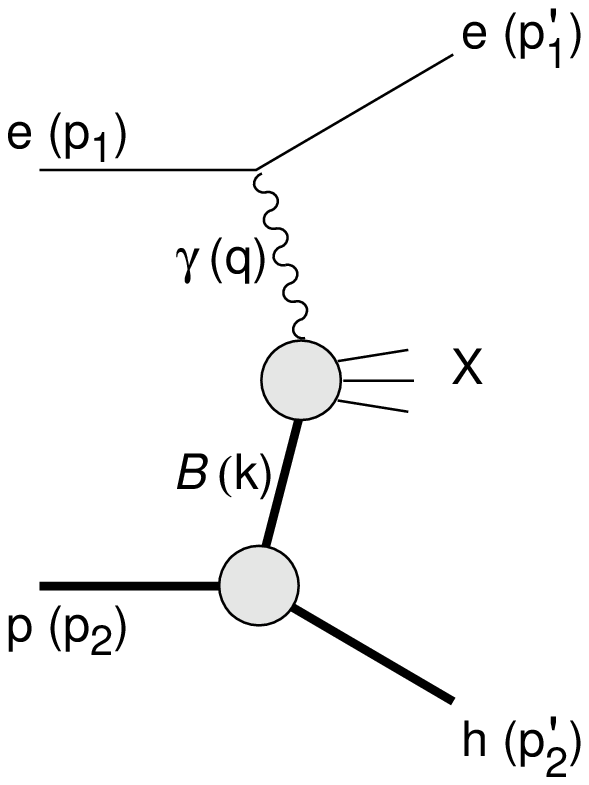}}}
\noindent{\eightrm Fig.11~~Single Reggeon exchange model of 
$\el\pr\ra\el{\rm h}X$.}
\vskip0.2cm
The moment of the polarised structure function $g_1^{\BB}$ of the 
Reggeon $\BB$ can be extracted from the polarisation asymmetry of the 
differential cross section moment in the limit $z\ra 1$:
\beqa
&&\int_0^{1-z} dx~ x~ {d\D\s^{target}\over dx dy dz dt} ~= \cr
&&{Y_P\over 2} {4\pi \alpha^2\over s} ~\Delta f(z,t)~
\int_0^1 dx_{\BB}~ g_1^{\BB}(x_{\BB},t;Q^2)
\eeqa
where $z=p_2^{\prime}.q/p_2.q$, $x_{\BB} = Q^2/2k.q$,
$1-z = x/x_{\BB}$, 
and $t = -k^2 \ll O(Q^2)$ so that $z\simeq\tilde z$.
The emission factor $\D f(z,t)$ cancels in the ratios.

Since the predictions (36) depend only on the $SU(3)$ properties
of $\BB$, together with target independence, they will hold equally 
well when $\BB$ is interpreted as a Reggeon rather than a pure hadron 
state. The ratios (36) can therefore be found (see e.g.~Fig.~12) by considering
the processes $\el\pr\ra\el\pi^-({\rm D}^-)X$ and $\el\ne\ra\el\pi^+
({\rm D}^0)X$.
\centerline{
{\epsfxsize=4.3cm\epsfbox{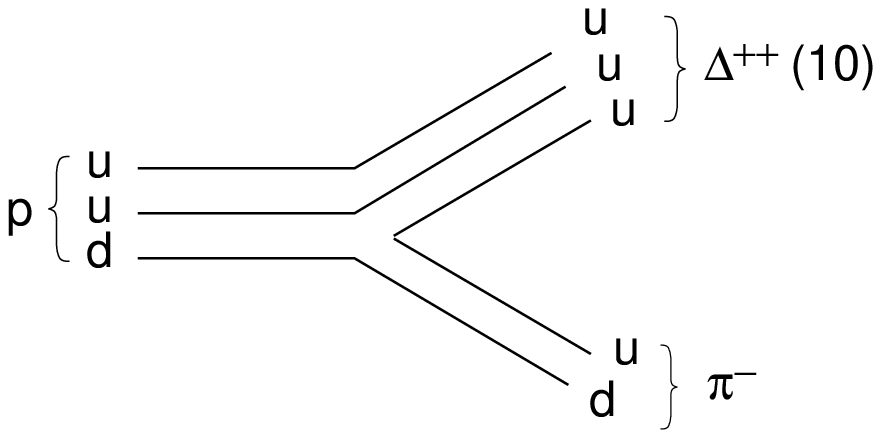}}}
\noindent{\eightrm Fig.12~~Quark diagram for the ${\rm Nh}{\cal B}$ vertex in
the reaction $\el\pr\ra\el\pi^- X$ where ${\cal B}$ has the quantum nos.~of
$\D^{++}$.}
\vskip0.2cm
Of course, the ratios (36) are only obtained in the limit as $z$ approaches 1, 
where the reaction $\el \Nu \ra \el \ha \Xx$ is dominated by the process
in which most of the target energy is carried through into the 
final state $\ha$ by a single quark. At the opposite extreme, for
$z$ approaching 0, the detected hadron carries only a small fraction of the 
target nucleon energy and has no special status compared to the other
inclusive hadrons $\Xx$. In this limit, the ratio of 
cross section moments for 
$\el \pr \ra \el \pi^-({\rm D}^-)\Xx$ and $\el \ne \ra \el \pi^+({\rm D}^0)\Xx$ 
is simply the ratio of the structure function moments for the proton and
neutron (35). 

Interpolating between these limits, we expect 
the ratios of  
$\int_0^{1-z} dx~ x~ {d\D\s^{target}\over dx dy dz dt}$
in the range $0<z<1$ for $\el \ne \ra \el \pi^+ ({\rm D}^0) \Xx$ 
over $\el \pr \ra \el \pi^- ({\rm D}^-) \Xx$ to look like the
sketch in Fig.~13.
\vskip0.2cm 
\centerline{
{\epsfxsize=6cm\epsfbox{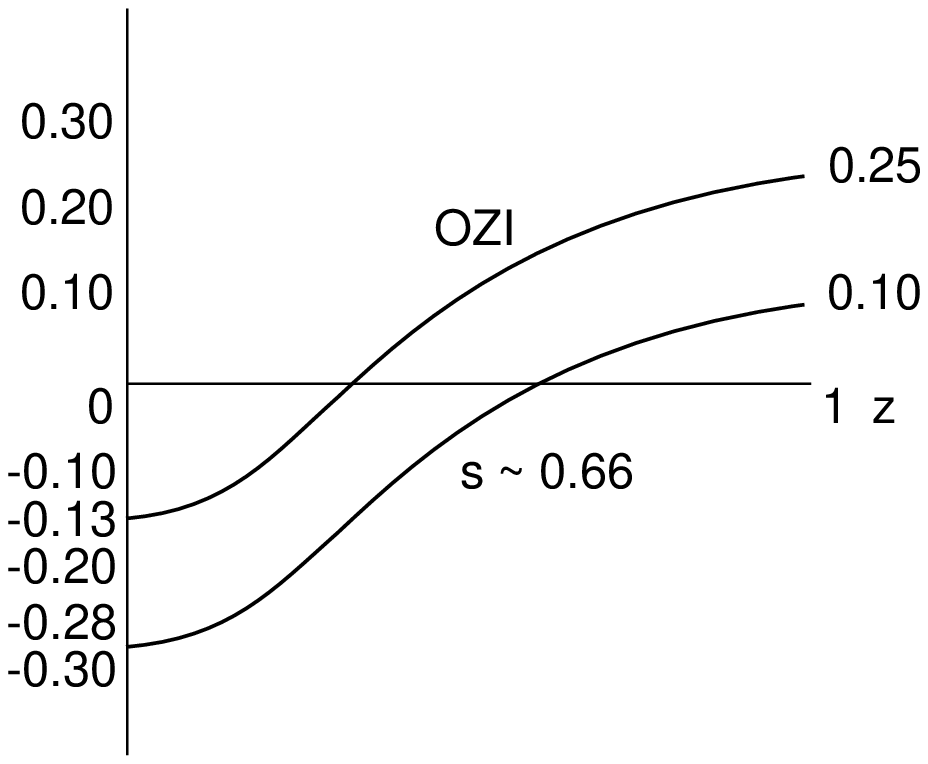}}}
\noindent{\eightrm Fig.13~~Cross section ratios for $\el\ne\ra\el\pi^+
({\rm D}^0)X$ over $\el\pr\ra\el\pi^-({\rm D}^-)X$ between $z\ra 0$ and $z\ra 1$,
contrasting the OZI and CPV predictions}
\vskip0.2cm
The difference between the OZI (or valence quark model) expectations 
and these predictions based on our target-independent interpretation
of the `proton spin' data is therefore quite dramatic, and should
give a clear experimental signal.

Since the proposed experiment requires particle identification in the
target fragmentation region, it is difficult to do at a polarised 
fixed-target experiment such as COMPASS\cite{COMP}, which is better suited
to studying semi-inclusive processes in the current fragmentation region.
A better option is a polarised $\el \pr$ collider, such as HERA\cite{HERA}. 
Testing our predictions requires comparision of proton and neutron data,
which can be extracted from experiments with polarised deuterons
replacing the protons in the collider.

\section{Acknowledgements}

I would like to thank S.~Narison and G.~Veneziano for their collaboration
on the original work described here, R.~Ball and S.~Bass for many helpful
discussions on the `proton spin' effect, and S.~Narison for once again
organising a stimulating and enjoyable Montpellier conference.

\end{document}